\documentclass[doublecol]{epl2}

\usepackage{graphicx}
\usepackage{color}

\title{Memory in random bouncing ball dynamics}
\shorttitle{Memory in random bouncing ball dynamics} 

\author{C. Zouabi\inst{1} \and J. Scheibert\inst{1} \and J. Perret-Liaudet\inst{1}}
\shortauthor{C. Zouabi \etal}

\institute{                    
  \inst{1} Laboratoire de Tribologie et Dynamique des Syst\`emes, UMR5513, CNRS/Ecole Centrale de Lyon/Univ Lyon/ENISE/ENTPE, 36 Avenue Guy de Collongue, F-69134 Ecully, France\\
}

\abstract{The bouncing of an inelastic ball on a vibrating plate is a popular model used in various fields, from granular gases to nanometer-sized mechanical contacts. For random plate motion, so far, the model has been studied using Poincar\'e maps in which the excitation by the plate at successive bounces is assumed to be a discrete Markovian (memoryless) process. Here, we investigate numerically the behaviour of the model for continuous random excitations with tunable correlation time. We show that the system dynamics are controlled by the ratio of the Markovian mean flight time of the ball and the mean time between successive peaks in the motion of the exciting plate. When this ratio, which depends on the bandwidth of the excitation signal, exceeds a certain value, the Markovian approach is appropriate; below, memory of preceding excitations arises, leading to a significant decrease of the jump duration; at the smallest values of the ratio, chattering occurs. Overall, our results open the way for uses of the model in the low excitation regime, which is still poorly understood.}

\begin{document}

\maketitle

\section{Introduction}
\begin{figure}[ht]
\includegraphics[width=\columnwidth]{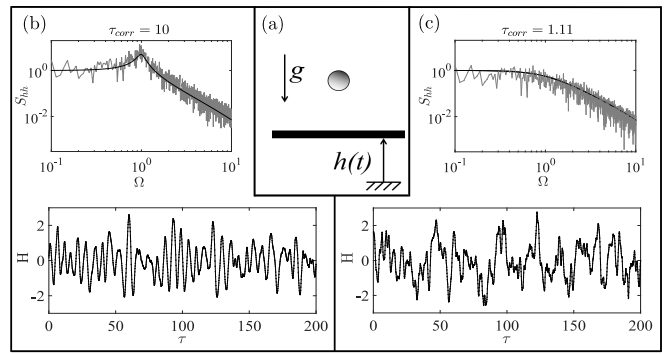}
\caption{(a) Sketch of the Bouncing Ball (BB) model. A ball submitted to gravity, $g$, bounces on an exciting plate having a correlated random displacement, $h(t)$. (b) (resp. (c)) Typical realisation of the dimensionless plate displacement, $H$, as a function of the dimensionless time, $\tau$ (bottom), and its power spectral density (PSD), $S_{hh}$ (top), for a dimensionless correlation time $\tau_{corr}$=10 (resp. 1.11). Solid lines are the analytic PSDs.\label{Fig1}}
\end{figure}

The bouncing ball (BB) model, \textit{i.e.} a point-like ball of finite mass bouncing vertically under the action of gravity, $g$, on an infinitely massive vibrating plate (Fig.~\ref{Fig1}(a)), has been widely studied in the last decades. Its popularity is due to both its simplicity and the richness of its dynamics, from harmonic to chaotic, through subharmonic and quasi-periodic solutions. It is now one of the paradigms for nonlinear dynamics and chaos \textcolor{black}{(see \textit{e.g.} \cite{Tufillaro1992, Guckenheimer1993} for BB in textbooks)}. It has been used to model a variety of systems, including granular flow~\cite{Linz1994,Ristow1994,Batrouni1996,Valance1998}, nanoscale mechanical contacts~\cite{Burnham1995}, bouncing droplets on a vibrating bath~\cite{Terwagne2013}, impact-induced noise~\cite{Kadmiri2012,Dang2013} and particle transport by surface waves~\cite{Ragulskis2008}.

In most cases, the BB model has been studied with a harmonic vertical motion of the plate (see \textit{e.g.}~\cite{Holmes1982, Luck1993, Giusepponi2005, Barroso2009, Vogel2011, Leonel2014}). \textcolor{black}{The BB model with random vibrations of the plate, in spite of its practical relevance to real systems and of the qualitatively new dynamic regimes it offers, has been much less investigated} (~\cite{Wood1981, Majumdar2007} for BB, or \cite{Karlis2006, DaCosta2015} for a modified BB model in which the ball is confined between two walls or when the ball is attached to a fixed wall through an elastic spring~\cite{DeAlcantaraBonfim2009}). In all these studies, the model was treated using a non-linear mapping (Poincar\'e map) for which the ball velocity $v_n$ just after the $n^{th}$ impact is related to $v_{n-1}$ and to the plate velocity $w_n$ at the instant of the $n^{th}$ impact. The randomness of the excitation was introduced by drawing each $w_n$ independently from a known distribution. The excitation process was thus assumed to be a memoryless, Markov chain.

Real vibration signals $w(t)$ are always characterized by a finite auto-correlation time $t_{corr}$ below which the signal keeps memory of its previous values. The limitation induced by the Markovian assumption of independent successive $w_n$ is thus qualitatively clear: the ballistic flight time of the ball between two successive bounces is assumed to be much larger than $t_{corr}$. This situation is more probable when the energy dissipation during impact is lower and/or when the characteristic plate velocity is higher. Note that for the so-called Chirikov conditions~\cite{Chirikov}, in which the height reached by the ball during a ballistic flight is large compared with the table displacement, a Markov-like excitation is expected. Conversely, for two bounces separated by a short flight-time, the two relevant plate velocities can be strongly correlated. Thus, in regimes in which short flight-times are dominant, the standard Markovian map approach is expected to fail to capture the BB model dynamics.

The purpose of this Letter is to characterize the BB model dynamics with stochastic excitation, when memory effects cannot be neglected. To do that, we performed full simulations of the ball dynamics using continuous excitations having various correlation times. We show how the statistical properties of the dynamics are affected by $t_{corr}$ and we identify the range of model parameters for which the Markovian map approach \textcolor{black}{is not valid anymore}.

\section{Model}
The originality of our model is to consider, for the vibrating plate, a correlated stochastic motion, $h(t)$, with tunable correlation time, $t_{corr}$. Figure~\ref{Fig1} shows typical realisations of $h(t)$, for either a long (b) or a short (c) $t_{corr}$. $h(t)$ is obtained from an (uncorrelated) Gaussian white noise, $\psi(t)$, filtered by a second order filter as:
\begin{equation}
\ddot{h}+2\zeta \Omega_c \dot{h}+\Omega_c^2 h=\psi(t),
\end{equation}
with $\Omega_c$ being the center frequency of the filter and $\zeta$ its damping coefficient. Note that $\zeta$ is related to the frequency contents of the signal, because the bandwith of its power spectrum density (PSD), $S_{hh}(\Omega)$, is equal to $2 \Omega_c \zeta$. $h(t)$ is twice differentiable in the least square sense, so we can define $w=\dot{h}$ and $a=\ddot{h}$. $h$ and $w$ have the property to be independent centered Gaussian random variables, with their stardard deviations being related by $\sigma_w = \Omega_c \sigma_h$. The autocorrelation function of $h$ is $<h(t)h(t+t')>=\sigma_h^2 e^{-\zeta \Omega_c \left|t'\right|} f(t')$, with $f$ a periodic function of $t'$~\cite{Preumont1994}, so that the correlation time can be defined as $t_{corr}=\frac{1}{\zeta \Omega_c}$. To avoid infinite energy in the acceleration signal, $a$, $h$ is further filtered by a first order low-pass filter with a cutoff frequency, $\Omega_L$, much larger than $\Omega_c$, so that this further filtering has negligible effect on $h$ and thus on $t_{corr}$ (in this study we use $\Omega_L=10\Omega_C$). The role of this extra filtering will appear later on. Typical PSDs of the signals used are shown in Fig.~\ref{Fig1}, together with their analytical prediction, for two cases: narrow-band (b) and broad-band (c).

For any generated excitation signal, $h$, we then solve the bouncing ball problem by calculating the values of the post-impact velocity, $v_n$, and instant, $t_n$, of all successive impacts. In practice, we solve the following equations:
\begin{eqnarray}
t_{n+1}=t_n +\theta_n, \label{time} \\
v_{n+1}=-e (v_n -g \theta_n) + (1+e) w_{n+1}, \label{velocity}
\end{eqnarray}
with $e$ being the restitution coefficient. The flight time, $\theta_n$, is obtained from the following equation:
\begin{equation}
-\frac{1}{2}g \theta_n^2 + v_n \theta_n + h_n - h_{n+1}=0. \label{flight}
\end{equation}

Equations \ref{time} and \ref{velocity} represent a classical Poincar\'e map for the BB model (see \textit{e.g.}~\cite{Guckenheimer1993}, section 2.4). Equation \ref{velocity} describes an instantaneous partially inelastic impact process, with energy dissipation quantified by $e\in[0;1]$ ($e$=1 and $e$=0 correspond to purely elastic and completely inelastic limit cases, respectively) and a pre-impact velocity $v_n -g \theta_n$ resulting from the finishing free flight. \textcolor{black}{Equation \ref{flight} enables analytic determination of the impact time, as the intersection between the parabolic free flight and the plate motion, $h$. In practice, $h$ is obtained using the fourth-order Runge-Kutta method with time step $\Delta t$ and piecewise linearized between integration points. To ensure that this interpolation is accurate, $\Delta t$ was chosen sufficiently small compared to the time period of the highest relevant frequency of the excitation signal, $2 \pi/\Omega_L$. As a consequence, the flight time determination was found to be essentially independent of the time step. Nevertheless, the shortest free flights are affected by the largest relative error on their flight time. This unavoidable limitation was handled by imposing} that the ball will stick to the plate if the estimate of the coming free flight duration, $\frac{2 v_n}{g}$, is found smaller than 10$\Delta t$, which amounts to define a cutoff velocity $v_{stick}=5g\Delta t$.
After sticking, the ball takes off again as soon as the acceleration of the plate becomes larger than gravity ($\ddot{h}>-g$); at this time, the ball takes off with an initial velocity equal to that of the plate.

Let us rewrite Eqs.~\ref{time} to~\ref{flight} in a dimensionless form:
\begin{eqnarray}
\tau_{n+1}=\tau_n+T_n, \label{timeadim} \\
V_{n+1}=-eV_n+\frac{e}{\Lambda}T_n+(1+e)W_{n+1}, \label{velocityadim} \\
-\frac{1}{2 \Lambda}T_n^2+V_n T_n + H_n-H_{n+1}=0, \label{flightadim}
\end{eqnarray}
with $\tau=\Omega_c t$, $T=\Omega_c \theta$, $H=\frac{h}{\sigma_h}$, $W=\frac{w}{\sigma_w}$, $V=\frac{v}{\sigma_w}$ and $\Lambda=\frac{\sigma_w^2}{g \sigma_h}$.
It is now clear that the system dynamics are fully controlled by two dimensionless parameters, the restitution coefficient, $e$, and the reduced plate acceleration, $\Lambda$. The effect of the correlation time, $t_{corr}$, will be obtained by varying $\zeta$, because the dimensionless form of $t_{corr}$ is $\tau_{corr}=\Omega_c t_{corr}=\frac{1}{\zeta}$. We have performed simulations for a large number of values of the triplet $\left\{\Lambda,e,\tau_{corr}\right\}$, with for each a large number of flights (the last $9.10^5$ out of $10^6$, to avoid any initial transient) so that accurate steady-state distributions of $V_n$ and $T_n$ could be obtained. In the following, $V_n$ will be denoted as the take-off velocity, because it represents initial flight velocities either after an impact or after a stick period. All simulations start with the ball lying on the plate and use $\Delta \tau=5.10^{-3}$.

\begin{figure}[h]
\includegraphics[width=\columnwidth]{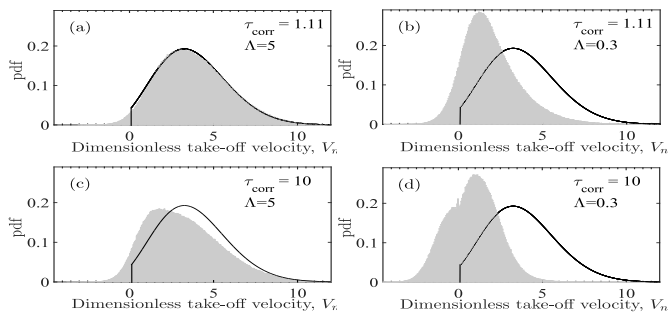}
\caption{Bars: Typical probability density functions (pdf) of the dimensionless take-off velocity, $V_n$, for selected $\Lambda$ and $\tau_{corr}$ (see legends), with the example of $e=0.8$. \textcolor{black}{Mean values are 3.65 in (a), 1.93 in (b), 3.12 in (c) and 0.85 in (d).} Solid lines: pdf of the fully uncorrelated case, $P_{WB}$ \textcolor{black}{(mean value 3.76)}.\label{Fig2}}
\end{figure}

\section{Results}
The case of a completely uncorrelated (Markovian) excitation has been studied by Wood and Byrne (WB)~\cite{Wood1981}. Assuming that (i) all impact velocities are downwards, (ii) all post-impact velocities are upwards and (iii) between impacts, reversal due to gravity always occurs, they provided the ($\Lambda$-independent) probability density function (pdf) of post-impact ball velocity, $P_{WB}(V)$, which depends \textcolor{black}{only} on $e$. We will use $P_{WB}$ as a reference to \textcolor{black}{highlight the differences brought by our improved model including correlation in the excitation}. Such memory effects are expected to be negligible when individual flight times are larger than the correlation time. This case is favoured for large excitations (large $\Lambda$) and/or short correlation time, $\tau_{corr}$. Figure~\ref{Fig2} indeed shows that, in our simulations, the pdf of $V_n$ closely matches $P_{WB}$ when $\Lambda$ is large and $\tau_{corr}$ is small (Fig.~\ref{Fig2}(a)). \textcolor{black}{Contrary to $P_{WB}$, our distributions are not only dependent on $e$, but also on the two other control parameters, $\Lambda$ (see Fig.~\ref{Fig2}(b)) and $\tau_{corr}$ (see Fig.~\ref{Fig2}(c). As expected, changes in the distribution occur when $\Lambda$ is reduced or when $\tau_{corr}$ is increased.} Combining a small $\Lambda$ and a large $\tau_{corr}$ produces even larger deviations from $P_{WB}$ (Fig.~\ref{Fig2}(d)). \textcolor{black}{Those deviations qualitatively change the distributions compared to $P_{WB}$. For instance, in Fig.~\ref{Fig2}(d), the mean value of the distribution  is 4.3 times smaller than that of $P_{WB}$, with almost half of the distribution corresponding to negative velocity values. Such negative velocities being forbidden in the WB analysis, $P_{WB}$ fails to capture, even qualitatively, the observed distributions in this range of control parameters.} Figure~\ref{Fig2} alone demonstrates that, although it has not been considered previously, the correlation time of the excitation has a significant influence on the BB dynamics.

\begin{figure}[ht]
\includegraphics[width=\columnwidth]{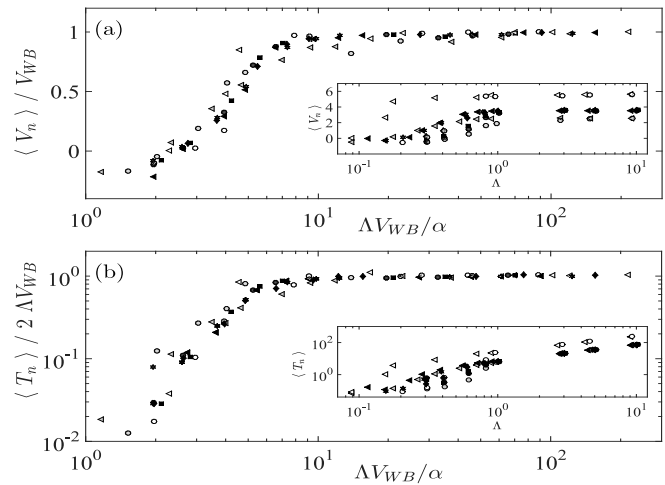}
\caption{(a) $\frac{\left\langle V_n\right\rangle}{V_{WB}}$ and (b) $\frac{<T_n>}{T_{WB}}$ as a function of $\frac{\Lambda V_{WB}}{\alpha}$, for various triplets $\left\{\Lambda, e, \tau_{corr}\right\}$. Grey, black and white symbols correspond to $e$=0.7, 0.8 and 0.9 respectively. Circles, squares, diamonds, stars and triangles correspond to $\tau_{corr}$=10, 8, 4, 2 and 1.11, respectively. Insets: (a) $<V_n>$ and (b) $<T_n>$ as a function of $\Lambda$ ; same data as in the main panels.\label{Fig3}}
\end{figure}

To quantify the effect of $\tau_{corr}$, let us now focus on the mean values of the distributions of the dimensionless take-off velocity and flight time, $<V_n>$ and $<T_n>$ respectively. The inset of Fig.~\ref{Fig3}(a) shows the evolution of $<V_n>$ as a function of the reduced acceleration, $\Lambda$, for various values of $\tau_{corr}$ and $e$. The main graph of Fig.~\ref{Fig3}(a) shows that all those velocity data can be collapsed onto a single master curve when plotting $\frac{<V_n>}{V_{WB}}$ as a function of $\frac{\Lambda V_{WB}}{\alpha}$, with $V_{WB}$ the mean value of $P_{WB}$ and $\alpha=\frac{\sigma_{w}^2}{\sigma_{h} \sigma_{a}}$, which is computed from the actual simulated excitation signal used. The physical interpretation of $\Lambda V_{WB}/\alpha$ will be discussed later on. $\alpha$ is a bandwidth parameter (see \textit{e.g.}~\cite{Preumont1994}, section 10.3) which depends only on the spectral contents of the excitation signal, \textit{i.e.} on $\tau_{corr}$ ; it is close to 1 (resp. 0) for a narrow-band (resp. wide-band) process. Note that $\sigma_{a}$ exists because of the $\Omega_L$-centered first order low pass filter. $V_{WB}$ depends only on $e$ and the function $V_{WB}(e)$ is provided in Fig.~11 of~\cite{Wood1981}.

An equally good collapse can be obtained with the time data when plotting $\frac{<T_n>}{T_{WB}}$ as a function of $\frac{\Lambda V_{WB}}{\alpha}$ (Fig.~\ref{Fig3}(b), main). $T_{WB}=2 \Lambda V_{WB}$ is the mean dimensionless flight time in the WB model, which straightforwardly follows from combining the classical relation $<T_n>=2 \Lambda <V_n>$ valid under Chirikov's assumption, with $<V_n>=V_{WB}$.

From Fig.~\ref{Fig3}, we can clearly define the range of parameters for which the BB dynamics are uncorrelated. It corresponds to the plateaus in both panels, \textit{i.e.} $\frac{\Lambda V_{WB}}{\alpha}$ larger than about 7. Indeed, in this regime, both $<V_n>$ and $<T_n>$ are equal to their uncorrelated counterparts, $V_{WB}$ and $T_{WB}$.
\textcolor{black}{Below 7, correlation effects are observed as both $\frac{<V_n>}{V_{WB}}$ and $\frac{<T_n>}{T_{WB}}$ become smaller than 1, as already illustrated for three particular cases in Fig.~\ref{Fig2}. The decrease is massive when $\frac{\Lambda V_{WB}}{\alpha}$ is decreased down to about 1: $\frac{<T_n>}{T_{WB}}$ drops by a factor as large as 100 ; concurrently $\frac{<V_n>}{V_{WB}}$ goes all the way down to zero and even changes sign.} 

\begin{figure}[ht]
\includegraphics[width=\columnwidth]{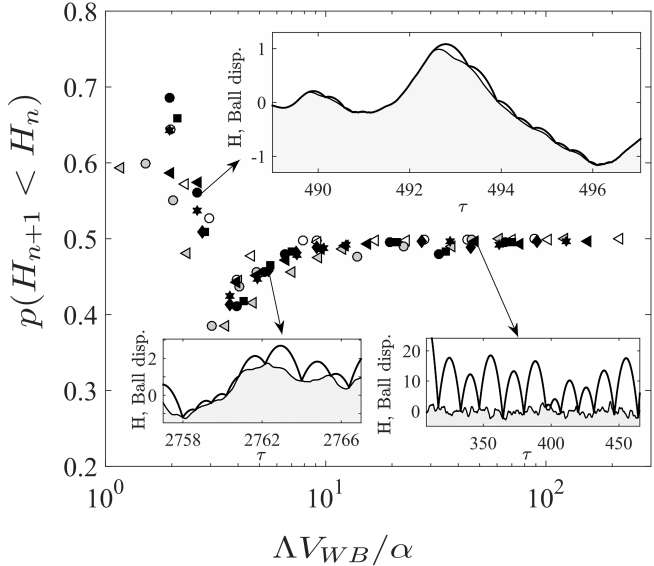}
\caption{Fraction of flights such that $H_{n+1}<H_n$, as a function of $\frac{\Lambda V_{WB}}{\alpha}$. Symbol colors and shapes have the same meaning as in Fig.~\ref{Fig3}. Insets: typical ball trajectories in three different simulations, each one being extracted from one of the three regimes described in the text. \label{Fig4}}
\end{figure}

In order to exhibit more clearly how the BB dynamics are correlated at low $\frac{\Lambda V_{WB}}{\alpha}$, we now consider the altitudes of two successive take-offs, $H_{n}$ and $H_{n+1}$. Figure~\ref{Fig4} shows the probability that $H_{n+1}$ be smaller than $H_{n}$, $p(H_{n+1}<H_n)$, for all performed simulations. We find that all data are reasonably collapsed onto a single master curve by the same dimensionless quantity, $\frac{\Lambda V_{WB}}{\alpha}$, that we already used in Fig.~\ref{Fig3}. Three regimes appear, the typical ball trajectories of which are illustrated as insets in Fig.~\ref{Fig4}:

- For $\frac{\Lambda V_{WB}}{\alpha}$ larger than about 7 (regime I), $p(H_{n+1}<H_n)$=50$\%$, which is consistent with fully uncorrelated dynamics: the ball can easily jump over any peak (local maximum) of the excitation $H(\tau)$ and thus has equal chances to impact the table again at a larger or smaller altitude.

- When $\frac{\Lambda V_{WB}}{\alpha}$ is decreased below 7 (regime II), an increasing fraction of jumps become smaller than the coming peak in $H(\tau)$. More and more jumps are required in order to climb the peaks. Helped by gravity, the way downhill is achieved in relatively less jumps. Overall, there are more jump uphill than downhill; which explains why $p(H_{n+1}<H_n)$ becomes smaller than 50$\%$.

- For values of $\frac{\Lambda V_{WB}}{\alpha}$ smaller than about 4 (regime III), another phenomenon is involved. Climbing the peaks requires so many inelastic jumps that their amplitude decreases down to the sticking limit. The rest of the climbing is thus achieved with the ball stuck to the plate, \textit{i.e.} with no more uphill jump. Take-off and jumps may start again when the plate goes down again, which explains why $p(H_{n+1}<H_n)$ rises, up to values exceeding 50$\%$.

- For $\frac{\Lambda V_{WB}}{\alpha}$ below  about 1, the probability of the ball to take-off when it is in a stuck state is almost zero. For instance, for $\frac{\Lambda V_{WB}}{\alpha} \simeq$ 1.42, we found less than 15 take-offs during a simulation time $\tau \simeq 6.10^6$. This explains why there is no point below 1 in Figs.~\ref{Fig3} and~\ref{Fig4}.

In regimes II and III, the free flights are significantly modified compared to the symmetric parabolae expected under Chirikov conditions, due to early or late intersections with the non-negligible table motion. This, or equivalently the fact that the term $H_n - H_{n+1}$ is not negligible anymore in Eq.~\ref{flightadim}, induces memory effects which translate into the drastic decrease of $<V_n>$ and $<T_n>$ observed in Fig.~\ref{Fig3} for $\frac{\Lambda V_{WB}}{\alpha}$ below 7.

\section{Discussion}
Given the above-described picture of the three bouncing regimes, we can now explain why $\frac{\Lambda V_{WB}}{\alpha}$ is the relevant parameter that controls the BB dynamics, including the emergence of memory effect. As we have seen, understanding the transition from uncorrelated (regime I) to correlated dynamics (regimes II and III) amounts to evaluating whether a single jump will be able or not to overcome the immediately coming peak (local maximum) of the plate motion. In other words, one needs to compare the typical duration of a jump with the typical time interval between two successive peaks. In dimensionless units, the former is directly $<T_n>$ which, at the transition from regime I to II, can still be approximated by $T_{WB}$ (Fig.~\ref{Fig3}(b)). The latter can be shown to be $T_{peak}=2 \pi \alpha$. The parameter used to rescale Figs.~\ref{Fig3} and ~\ref{Fig4} is thus naturally proportional to the ratio $\frac{T_{WB}}{T_{peak}}$ between these two characteristic time scales.

\textcolor{black}{We emphasize that $\frac{\Lambda V_{WB}}{\alpha}$ combines into a single quantity all three control parameters of the model: $\Lambda$, $e$ (through $V_{WB}$) and $\tau_{corr}$ (through $\alpha$). This finding will considerably simplify further studies of the correlated stochastic BB model by drastically reducing the dimension of the parameter space to be explored, from three down to only one.} 

To complete the story about memory effects in the BB dynamics caused by correlation in the plate motion, we now need to identify the relationship between $\alpha$ and $\tau_{corr}$. There is no universal such relationship, because $\alpha$ depends on the shape of the PSD of $h(t)$. In our case of a white noise filtered by a second order filter and extra-filtered by a first order filter with a cut-off frequency $\Omega_L$, one can show that $\alpha(\tau_{corr})=\left[\left(1+\frac{2 \Omega_L}{ \Omega_c \tau_{corr}}\right) \left(1+\frac{2 \Omega_c}{ \Omega_L\tau_{corr}} \right)\right]^{-1/2}$. We found that this formula is in perfect agreement with calculations of $\alpha$ done using the various generated $h(t)$.

Regime III bears many analogies with the important phenomenon of chattering (or inelastic collapse), which consists in the sticking of the inelastic ball onto the plate after a finite time, but an infinite number of bounces. Again, so far, this phenomenon has been studied for periodically vibrating tables (see \textit{e.g.}~\cite{Luck1993,Budd1994,Giusepponi2003}) or with Markovian random excitations~\cite{Majumdar2007}. Here, we also observe chattering in the case of correlated random excitations: regime III is a random succession of bouncing and sticking periods, the statistics of which are found to strongly depend on the value of the correlation parameter $\frac{\Lambda V_{WB}}{\alpha}$ (see the strong variations of $p(H_{n+1}<H_n)$ with $\frac{\Lambda V_{WB}}{\alpha}$ when the latter becomes smaller than about 4).

The transition from regime I (uncorrelated) to regime II (correlated bouncing dynamics) is expected to be independent of the details of our model, in particular of the precise PSD used. In contrast, the transition from regime II to regime III (correlated random chattering) may depend on modelling choices. In our simulations, the onset of sticking depends on an arbitrary cutoff velocity $v_{stick}$. Although we have found that doubling and halving $v_{stick}$ has negligible influence on our results, the sticking process is not identical to the pure inelastic collapse case, in which the ball velocity goes down to arbitrarily small values before sticking. We believe that extending our model to include pure collapse conditions will provide significant insights into random chattering.

Overall, our results provide new insights into how and why memory arises in BB dynamics when realistic, continuous correlated excitations are concerned. The main effect of memory is to significantly decrease the jump duration with respect to its expected value for uncorrelated excitation. The present study opens new research avenues to apply the BB model to low excitation levels or to understand the transition from sticking to bouncing dynamics when the excitation is increased. Our work is part of the effort to include memory effects in some of the most prominent non-linear dynamics and statistical physics models, including the BB (this work), the random walk~\cite{Guerin2016}, the predator/prey~\cite{Aly2006}, the foraging~\cite{Kerster2016} and the Ising~\cite{Brey2002,Caccioli2008} models.

\acknowledgments
This work was supported by the LABEX MANUTECH-SISE (ANR-10-LABX-0075) of Université de Lyon, within the program "Investissements d'Avenir" (ANR-11-IDEX-0007) operated by the French National Research Agency (ANR). The research leading to these results has received funding from the People Programme (Marie Curie Actions) of the European Union's Seventh Framework Programme (FP7/2007-2013) under REA grant agreement n° PCIG-GA-2011-303871. We thank A. Le Bot, G. Pallares and D. Dalmas for their comments on the manuscript.



%

%
%
%
%
%



%
%
%
%


\end{document}